# An operational framework to automatically evaluate the quality of weather observations from third-party stations

Quanxi Shao*, Ming Li, Joel Janek Dabrowski, Shuvo Bakar, Ashfaqur Rahman, Andrea Powell, and Brent Henderson

Data61, CSIRO, Australia
`Firstname.lastname@data61.csiro.au`

**Abstract.** With increasing number of crowdsourced private automatic weather stations (called TPAWS) established to fill the gap of official network and obtain local weather information for various purposes, the data quality is a major concern in promoting their usage. Proper quality control and assessment are necessary to reach mutual agreement on the TPAWS observations. To derive near real-time assessment for operational system, we propose a simple, scalable and interpretable framework based on AI/Stats/ML models. The framework constructs separate models for individual data from official sources and then provides the final assessment by fusing the individual models. The performance of our proposed framework is evaluated by synthetic data and demonstrated by applying it to a real TPAWS network.

## 1 Introduction

Observations are a key to understand our surrounding environment, particularly in the changing climate we are facing, in which local observations are necessary to capture small scale variations and rare events. With the technical advancement and cost reduction, more and more private bodies, organizations and local governments install crowdsourced weather stations (called TPAWS – Trusted Private Automatic Weather Station) in their premises for different purposes. For example, the department of primary industries and regional development (DPIRD[1]) of Western Australia has built a network of automatic weather stations throughout the state, which provides timely, relevant and local weather data to assist growers and regional communities to make more-informed decisions. Goanna Ag[2] has created a comprehensive weather station and soil moisture probe network to assist its smarter farming business.

While these TPAWS observations may provide extra information, its quality is a major concern [3]. For example, farms can use their local weather stations to help its management and assessment of extreme weather events (e.g., frosts and heatwaves) and as evidence to make insurance claims, but other users such as the Agricultural Finance/insurance industry need to have confidence in the reported local observations. In order to incorporate the TPAWS observations in policy criteria and to verify/assess claims, it is necessary to cross check the reliability of the reported local observations with observations from reliable sources such as the high-quality weather stations from the Australian Bureau of Meteorology (BoM) and the coarser endorsed products (such as ACCESS NWP gridded data product[4]). From now on, the data used as the reliable sources are called the official data. The reliability of the reported TPAWS observations can also be checked by the consistency of its own observation series over time.

From the viewpoint of Ag Finance / insurance industry, if they can obtain the confidence and are able to assure the quality of the reported local observations, it may allow them to use the observations for its policy and product design across these industrial sectors and to deliver more tailored insurance products to these sectors (e.g., reduce premiums in some cases, get more farmers onboard as clients, and reduce the disagreements on relative claim). For farmers it would enable a better representation of weather events and thus set a more appropriate assessment of a claim by the insurer.

To develop an operational framework for targeting the need from Agricultural Finance / insurance industry as well as farmers, CSIRO Data61 entered into an agreement with BoM, to develop a set of AI/Statistical/Machine Learning models, aiming to (a) check whether TPAWS observations have been tampered with or caused by equipment failure; and (b) estimate degree of confidence of observations submitted by TPAWS network. The weather variables most of interest are rainfall (Rain) and minimum temperature (Tmin), followed by maximum temperature (Tmax), wind and relative humidity. We present the framework in this paper along with brief descriptions of AI/Stat/ML methods, which are simple, scalable and interpretable in the setting for operational purpose but powerful and robust to achieve high level of assessment results. The novelty of this work lies in the application space where quality of TPAWS observation is assessed by fusing a set of heterogeneous data sources (remotely sensed and ground based) based on a set of complementary AI/Stat/ML methods (spatial, temporal, and spatiotemporal).

The paper is organized as following: the potential data sources used in the paper is given in Section 2, followed by the proposed framework in Section 3, and evaluation using synthetic wind gust data and application to DPIRD maximum daily temperature data in Section 4. Conclusions and a discussion are given in Section 5.

## 2 Official Data

The following data are used as official data sources in assessing TPAWS observations:

*BoM station observations*: which refer to high-quality daily weather observations from BoM networks and provide the most accurate weather measurements after appropriate quality control. The BoM networks consist of ~700 automatic weather stations cross the Australian continent, generating and handling data automatically. The BoM weather observations are available from the BoM Climate Data Online (BoM[5]).

*ACCESS NWP forecasts*: The Australian Community Climate and Earth-System Simulator (ACCESS) Numerical Weather Prediction (NWP)[6] data are made by operational NWP systems adopted by the BoM. Weather forecasts at hourly intervals up to 240 hours are issued by the ACCESS systems four times daily (00:00, 06:00, 12:00 and 18:00 UTC) with ~12km resolution. The ACCESS NWP APS2 forecast data are available from 07/06/2016 to 25/09/2020 and can be directly downloaded from National Computational Infrastructure (NCI) with an OpenDAP server option.

*AGCD data*: The Australian Gridded Climate Data (AGCD) is the BoM's official dataset for climate analyses covering the weather variables of rainfall, maximum temperature, minimum temperature and vapour pressure at daily and monthly time-scales[7]. The AGCD of daily weather variables with 5km resolution is available from 1900 onwards. The AGCD can be obtained directly from the BoM Climate Data Online[5].

*ERA data:* ECMWF atmospheric reanalysis (ERA), which is the global climate and weather reanalysis data which covers the period from 1979 to present and provides hourly wind gust on a 30km grid [8]. ERA data is available from the NCI data collection.

*Radar data*: Rainfields is the BoM current system for quantitative radar estimation for real-time, spatially and temporally continuous rainfall data [9]. There are a total of 63 rain radars available as of 2022 in Australia [10]. Rainfields data are available within a radius up to 256 km from a radar and at 1 km resolution with 5-minute updates. Rainfields data can be requested directly from the BoM Climate Data Online[4].

Different official data sources are used for different weather variables and different tests (**Table 1**).

**Table 1.** The application of official data sources to weather variable and tests

| Official data sources | Weather variables | Tests |
|---|---|---|
| BoM station observations | Tmax, Tmin, Wind, Humidity | Spatial, Spatial temporal, Trend |
| ACCESS NWP forecasts | Tmax, Tmin, Wind, Humidity | NWP |
| AGCD data | Tmax, Tmin, Rain, Humidity | AGCD |
| ERA data | Wind | ERA |
| Radar data | Rain | Radar |

## 3 Proposed Framework

Several issues need to be addressed in the operational framework. Firstly, there is currently a manual process in the Bureau's QC procedure with assistance from a series of statistical tests for the data from Bureau's own weather stations. The high-level challenge is to make the processes automatic as required as a framework. Secondly, the BoM's QC procedure is undertaken under the assumption that the data should be accepted if there are no clear reasons to reject the observations. However, this assumption is invalid for assessing TPAWS observations. Thirdly, the current BoM's QC procedure reports a "Flag" if an observation is suspected, while the framework for assessing the TPAWS observations needs to provide as a confidence level. Finally, there are many manual stations in the BoM's network, which cannot be used in the assessment because the framework needs to deliver assessment result in near-real time (e.g., the next day) before the data from manual stations become available, making the number of stations used in the assessment smaller (only around 700 stations in total).

In term of methodological development, this framework is to some extent an attempt to automate the assessment procedure but also provide a transparent and scalable verification approaches underpinning the procedure. It is important for the framework to be as simple as possible and widely applicable to different weather variables as it will need to be operational on the BoM system. Furthermore, the designed framework should be flexible to handle different number of official data sources (e.g., including new sources of trusted data once they are available and/or excluding existing source of trusted data when they are reported as unavailable /missing). Finally, the assessment results should be interpretable in the sense that users can trace back where the disagreement is between TPAWS observations and the trusted data sources. To satisfy all these requirements, our framework is designed as hierarchical structure through (1) constructing separate models against to individual official data sources and (2) combining the individual assessment results together to derive an overall assessment. To avoid the duplication of information use, the framework will only select the best individual model for a particular official data source. Note that the TPAWS observations are subject to errors. Robust treatment should be applied in the model calibration. A domain test is always applied to identify obvious error in a TPAWS observation before further assessment. **Fig. 1** presents the framework for assessing maximum daily temperature (Tmax) as an example. More details on individual tests and final assessment are given below.

*Domain test*: before carrying out any further tests, a domain test is performed to check whether the value of a TPAWS observation falls within physical limits (**Table 2**). A domain test can quickly identify obviously erroneous observations.

*Spatial test*: A spatial test checks a TPAWS observation against a spatial prediction from nearby BoM observations. The spatial prediction is made by linearly combining the correlation of target TPAWS observations with nearby BoM station and the linear coefficients are estimated by the least absolute shrinkage and selection operator (LASSO[11]) method to facilitate robust parameter estimation. The uncertainty of the spatial prediction is quantified based on an error model to characterize the difference between observations and the corresponding spatial predictions. The log-sinh transformation[12] is applied to both observations and predictions for data normalization and variance stabilization. The spatial test requires at 2 BoM stations available within a 200km radium.

*Spatial temporal test*: Similar to the spatial test, a spatial test checks a TPAWS observation against a spatial-temporal prediction from current and past observations of nearby BoM stations. The algorithm exploits screening processes through statistical techniques, such as Hampel filter, trend analysis, analysis of variance (ANOVA), Tukey's Honest Significant Difference (HSD) test and LASSO to identify similarities between stations. After identifying the similar stations, we perform three different types of spatial-temporal models (i) STAR (spatial-temporal autoregressive) (ii) STAM (spatial-temporal additive model) and (iii) STLM (spatial-temporal linear model) to get prediction at the TPAWS site[13]. Finally, ensemble Bayesian model averaging method is used to obtain the weighted spatial-temporal prediction with corresponding uncertainty.

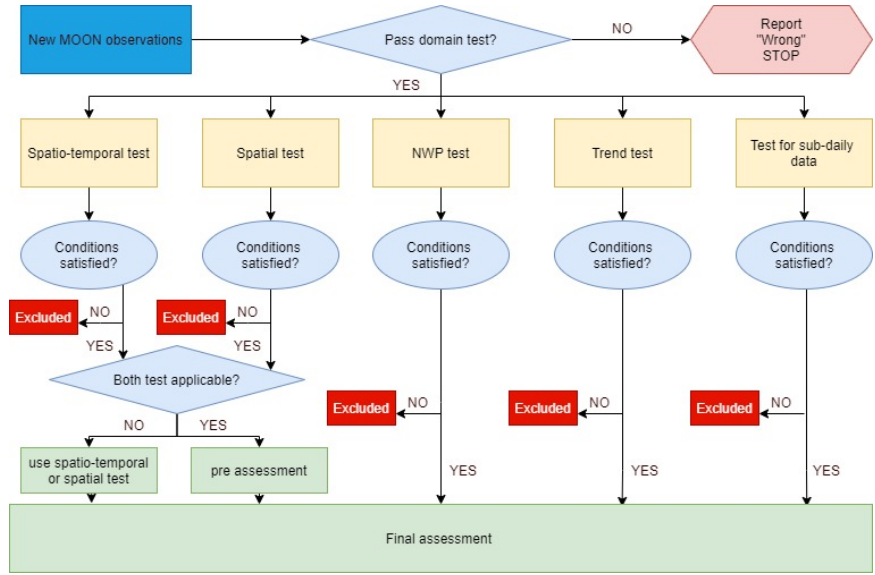

**Fig. 1.** A schematic flowchart outlining the framework for assessing the TPAWS observations. The framework includes a set of tests to assess the quality of TPAWS observations from different perspectives and an algorithm to merge the results from appropriate tests to provide a unified final assessment.

**Table 2.** The physical limits of weather observations used in the domain test for Australia

| Weather variables | Lower limits | Upper limits |
|---|---|---|
| Tmax | Tmin | 60℃ |
| Tmin | −30℃ (or −40℃ if above 1000m) | Min(60℃, Tmax) |
| Rain | 0 | 2000 mm |
| Wind | 3.6 km/h | 540 km/h |
| Humidity | 0% | 100% |

*Trend test*: A trend test is designed to check observations against previous day's observation with consideration of the change estimated from nearby BoM stations. The trend test is equivalent to the spatial test applied to the change of weather observations in two consecutive days.

*Gridded data test*: which compares the difference between TPAWS observations and gridded data (such as ACCESS NWP, AGCD, ERA and radar data). Same as the spatial test, the gridded data test uses an error model to estimate the uncertainty of grided data as a prediction.

*Subdaily test*: which fuses data from a TPAWS observation and a BoM grid data product to make predictions over time. This is performed using a Dynamic Linear

Model (DLM) [14] together with the Kalman Filter. The DLM fuses the TPAWS observation and grid data and models the dynamics of a target weather variable (e.g., temperature) over a 24-hour period. The Kalman filter is used to perform inference in the DLM model by updating the predictions at each time given the past data. Each prediction in time is in the form of a Gaussian distribution with a mean and variance. For example, in the case of Tmax, a recorded daily Tmax can be evaluated according to the predicted temperature distribution. If it does not fit the distribution well, it will have a low confidence.

Each individual test provides a probabilistic prediction of the target variable $X$ based on different information. In this study, the confidence level (CL) of a measurement $x$ is measured in the framework of hypothesis tests by a two-sided p-value based on the $i$-th method $F_i$, denoted by $p_2(x|F_i)$, as follows:

$$CL = p_2(x|F_i) = 1 - 2|p_1(x|F_i) - 0.5| \qquad (1)$$

where $p_1(x|F_i) = P(X \leq x)$ is the one-sided (left) p-value based on $F_i$. A greater value of CL indicates a more reliable measurement. If the measurement is close to the prediction from a test method, CL will be close to 100%. Conversely, a measurement far from the prediction leads to CL close to 0%.

The final assessment on TPAWS observations with the basic logic is given as below (**Fig. 1**):

(1)    TPAWS observations are firstly checked by a domain test to see whether the value is physically meaningful.

(2)    After passing domain tests, five individual tests (including spatial-temporal test, spatial test, NWP test, trend test and test for sub-daily data) are carried out parallelly to test the quality of TPAWS observations from different aspects.

(3)    Conditions associated with each test will be checked to determine if the test can be used in the final assessment.

(4)    The final assessment is derived by fusing the results from the tests that satisfy assumptions and conditions.

It should be noted that when both spatial-temporal test and spatial test are applicable, a pre-assessment is used to determine which of these two tests will be used in the final assessment. The pre-assessment calculates the optimal weights of different tests if either spatial-temporal test or spatial test is used together with other applicable tests. The test (e.g., spatial-temporal test or spatial test) with a greater weight will be considered as the better test and will be used in the final assessment and the other one will be excluded to avoid the duplication of information usage. The final assessment weights the results from different tests and provides an overall confidence level for the TPAWS observations. In case that all tests are not applicable, the final assessment will not be delivered and a "NA" (Not Applicable) will be provided. As a low confidence in the overall assessment is caused by low confidences in one or more tests, a user can trace back which test(s) provide low confidence in the case when the overall assessment indicate a lower confidence.

## 4    Performance Evaluation and Application to DPIRD Data

To evaluate the performance of our proposed framework, synthetic dataset was generated at selected 100 BoM stations in 2016-2019 by the BoM technical team. We only use wind gust as an example here. In a nutshell, small random but positive errors in a range from 5m/s to 14.6m/s were added to ~10% actual wind gust observations. Model evaluation is based on the data in 2018-2019 and model training is based on the data in 2016-2017. Table 3 presents the summary statistics for the spatial, ERA, NWP tests and the merged test by combining these three tests. As expected, the merged test performs slightly worse for extremely calm wind and extremely strong wind than medium wind. Furthermore, our test tends to be less skillful at stations either at a remote place, on the costal line or at high altitude.

**Table 3.** The summary statistics for the spatial, ERA, NWP tests and the merged test by combining these three tests

| | | Test methods | | | |
| --- | --- | --- | --- | --- | --- |
| | | Spatial | ERA | NWP | Merged |
| Hit rate (%) | All | 92.8 | 82.8 | 82.0 | 95.3 |
| | <25km/h | 99.2 | 92.6 | 94.3 | 98.5 |
| | 25-60 km/h | 92.9 | 83.0 | 81.5 | 95.4 |
| | >60 km/h | 93.9 | 68.9 | 75.1 | 89.9 |
| False alarm rate (%) | All | 5.0 | 4.3 | 4.2 | 4.5 |
| | <25km/h | 7.6 | 4.2 | 3.9 | 6.6 |
| | 25-60 km/h | 4.5 | 3.7 | 3.6 | 3.8 |
| | >60 km/h | 7.3 | 9.4 | 9.4 | 10.2 |

We also applied the framework to evaluate DPIRD daily Tmax data. Fig. 2 presents the cumulative distribution of the proportion of suspect observations at each DPIRD station. We can see that only few suspect observations have been identified by the framework at most stations. Fig. 2 also shows an example of suspect DPIRD observations. This example is taken from a DPIRD station (DU002) located at Dumbleyung. The Tmax measurement on 03/Oct/2019 is 45.8 ℃ and looks much larger than the measurements from adjacent days. The confidence level of this observation is 0%, confirming a highly suspect measurement.

## 5    Conclusions and a Discussion

Aiming to simplicity, scalability and interpretability for operational system to assess observations from rapidly growing TPAWS networks, we developed a framework based on a set of AI/Stat/ML methods, to produce overall level of confidence by fusing a set of fast and robust assessment models for individual official data sources. The framework is flexible enough to handle any possible individual models through intermediate selection for the models using the same official data sources. Individual models are developed for all the variables of interest including the maximum and minimum daily temperatures, daily total rainfall, daily wide gusts and humidity at 9:00 am and 3:00pm. The performance of our designed framework is tested by using synthetic data based on randomly selected data from the Bureau's official weather stations and demonstrated by applying the framework to DPIRD's AWS. The framework has been implemented in BoM operational system as a potential product which is being tested on a number of crowdsourced TPAWS networks.

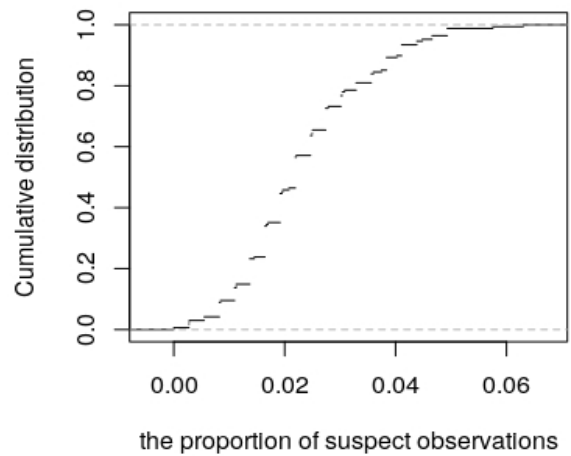

| Date | Min temp °C | Max temp °C | Min RH % | Max RH % | Rain mm |
|---|---|---|---|---|---|
| 05-10-2019 | 12.7 | 19.5 | 47 | 99.8 | 5.8 |
| 04-10-2019 | 13.8 | 23.3 | 31.2 | 86.6 | 0 |
| 03-10-2019 | 8.5 | 45.8 | 1 | 99.8 | 0 |
| 02-10-2019 | 11.5 | 22.3 | 34.6 | 99.7 | 0 |
| 01-10-2019 | 10.3 | 23.8 | 32 | 99.8 | 0 |

**Fig. 2.** (top) The cumulative distribution of the proportion of suspect Tmax observations at each DPIRD station and (bottom) an example of suspect DPIRD observations.

The underlying assumption in this framework is that the TPAWS observations are partially trusted in the sense that their observations are overall reliable (otherwise, all the designed models would be questionable) so that the models can be calibrated.

Therefore, a quality assurance procedure should be taken before conducting the assessment. The framework is not applicable to the case all observations used for model calibration are manipulated (e.g., deliberately by malicious parties).

Furthermore, the length of observations from each TPAWS site should be enough to calibrate the models. Throughout our tests, we found that there should be at least 2 years of daily observations for the calibration. We also considered the use of satellite data but the results showed that they did not contribute to any value to the final assessment. Our framework is flexible to include more official data easily once they are available. The framework can be potentially applied to automate BoM QC procedure and is applicable to other countries where different official data sources are used. To the best of our knowledge, our framework is the first attempt to automatically assess unofficial crowdsource data by using official data sources.

## Acknowledgement:

We thank the BoM for their assistance in providing the data (including the synthetic data) and for the many discussions throughout the technical development. Constructive comments and suggestions by two anonymous reviewers are also acknowledged.